# Measuring Two-Event Structural Correlations on Graphs

Ziyu Guan
Dept. of Computer Science
University of California
Santa Barbara, CA 93106, USA
ziyuguan@cs.ucsb.edu

Xifeng Yan
Dept. of Computer Science
University of California
Santa Barbara, CA 93106, USA
xyan@cs.ucsb.edu

Lance M. Kaplan
U.S. Army Research Laboratory
Adelphi, MD 20783 USA
lance.m.kaplan@us.army.mil

## ABSTRACT

Real-life graphs usually have various kinds of events happening on them, e.g., product purchases in online social networks and intrusion alerts in computer networks. The occurrences of events on the same graph could be correlated, exhibiting either attraction or repulsion. Such structural correlations can reveal important relationships between different events. Unfortunately, correlation relationships on graph structures are not well studied and cannot be captured by traditional measures.

In this work, we design a novel measure for assessing two-event structural correlations on graphs. Given the occurrences of two events, we choose uniformly a sample of "reference nodes" from the vicinity of all event nodes and employ the Kendall's $\tau$ rank correlation measure to compute the average concordance of event density changes. Significance can be efficiently assessed by $\tau$'s nice property of being asymptotically normal under the null hypothesis. In order to compute the measure in large scale networks, we develop a scalable framework using different sampling strategies. The complexity of these strategies is analyzed. Experiments on real graph datasets with both synthetic and real events demonstrate that the proposed framework is not only efficacious, but also efficient and scalable.

## 1. INTRODUCTION

In recent years, an increasing number of real-life networks have emerged and experienced a substantial growth, e.g. online social networks, WWW and Internet. A lot of works have been dedicated to research problems related to network structures, e.g. graph pattern mining [17, 25] and link analysis [6]. One important aspect of complex networks is that their nodes usually produce various kinds of data, which we abstract as events in this work. For instance, an eBay customer could sell or bid a product; a computer in Internet could suffer various attacks from hackers. This gives birth to the branch of research involving graph structures and events happening on graphs [8, 20, 16, 11].

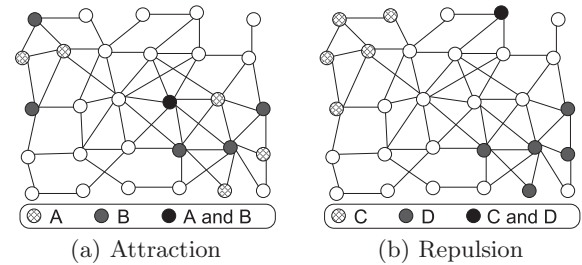

**Figure 1: Two types of Two-Event Structural Correlation: (a) attraction and (b) repulsion.**

Two events occurring on the same graph could be correlated. Two illustrative examples are shown in Figure 1. In Figure 1(a), A and B exhibit a *positive correlation* (attraction). In the context of a social network, they could be two baby formula brands, Similac and Enfamil. Their distributions could imply that there exist "mother communities" in the social network where different mothers would prefer different baby formula brands. The two brands attract each other because of the communities. An example of *negative correlation* (repulsion) could be that people in an Apple fans' community would probably not buy products of ThinkPad and visa versa, as conveyed by Figure 1(b). We name this kind of structural correlation as *Two-Event Structural Correlation* (TESC). TESC is different from correlation in transactions such as market baskets. If we treat nodes of a graph as transactions and assess Transaction Correlation (TC) of two events by using measures such as Lift [12], one can verify that in Figure 1(a), A and B have a negative TC, although they exhibit a positive TESC. Regarding the baby formula example, a mother would probably stick to one brand, since switching between different brands could lead to baby diarrhea. As another example, in terms of computer networks, A and B could be two related intrusion techniques used by hackers to attack target subnets. Since attacks consume bandwidth, there is a tradeoff between the number of hosts attacked and the number of techniques applied to one host. Hackers might choose to maximize coverage by alternating related intrusion techniques for hosts in a subnet, in order to increase the chance of success. We will show such examples in the experiments. Hence, TESC is useful for detecting structural correlations which might not be detected by TC. It can be used to improve applications such as online advertisement [4] and recommendation [14]. For instance, most recommendation methods exploit posi-tive TC, while positive TESC provides an alternative recommendation scheme in local neighborhoods. TESC can reveal important relationships between events (the intrusion example) or reflect structural characteristics (communities in the product examples) on a graph. Nevertheless, this paper focuses on measuring TESC, but not finding its cause.

Unfortunately, measuring TESC is not a trivial problem. A similar problem also exists in correlation test between two point sets in spatial data [7, 18, 23]. However, discrete graph space is intrinsically different from continuous spatial space and the existing techniques for the point pattern problem cannot be applied directly. More importantly, in the graph context, scalability is a major issue since real-life networks often contain millions or even billions of nodes and edges, while the point pattern problem only considers datasets of very limited sizes (e.g. several hundreds of points).

Researchers have studied the distributions of events in graph spaces. Kahn *et al.* studied proximity pattern mining where the goal was to mine groups of events which frequently co-occurred in local neighborhoods in a graph [16]. However, (1) they only consider positive correlations (association) among events, while TESC aims to measure both positive and negative correlations; (2) they rely on an empirical method for significance testing, while our method is rigorous statistical testing based on the Kendall's $\tau$ statistic; (3) their problem is intrinsically a frequent pattern mining problem and could miss some rare but positively correlated event pairs, as will be shown in the experiments. In [11], we proposed a measure based on hitting time to assess the structural correlation within an event. That measure is not suitable for TESC in that if we adapt the measure to compute the affinity between two events, its distribution in the null case is difficult to estimate by simulations. It is hard to preserve each event's internal structure when simulating independence between them.

In this paper, we propose a novel measure and then an efficient framework for computing TESC on graphs. Specifically, given the occurrences of two events, we choose a sample of "reference nodes" uniformly from the vicinity of all occurrences and compute for each reference node the densities of the two events in its vicinity, respectively. Then we employ the Kendall's $\tau$ rank correlation measure [15] to compute the average concordance of density changes for the two events, over all pairs of reference nodes. Finally, correlation significance can be efficiently assessed by $\tau$'s nice property of being asymptotically normal under the null hypothesis. For efficiently sampling reference nodes, different sampling techniques are proposed to shorten the statistical testing time. Our framework is scalable to very large graphs.

**Our Contributions** We introduce a new structural correlation problem: Two-Event Structural Correlation (TESC), which measures whether and to what degree two events occurring on the same graph are correlated. TESC is a fundamental problem that helps understand how different events are related with one another on a particular graph and can be insightful for many attributed graph mining applications. Second, we develop an efficient statistical testing framework for measuring TESC. The main idea is to compute the concordance of each pair of reference nodes with regard to the respective density changes of the two events, from one reference node's vicinity to the other's. High concordance scores mean that the occurrence of one event tends to attract the occurrence of the other event (positive correlation), while low concordance scores mean the two events repulse each other (negative correlation). An important subproblem is how to efficiently sample reference nodes. To this end, we propose three different algorithms for efficient reference node selection and analyze their advantages and disadvantages both theoretically and empirically. Finally, we demonstrate the efficacy of the TESC testing framework by event simulations on the DBLP graph. We further test its scalability in a Twitter graph with 20 million nodes. Case studies of applying the testing framework on real events occurring on real graphs are provided with interesting results.

## 2. PRELIMINARIES

We consider an attributed graph $G = (V, E)$ where an event set $Q$ contains all events that occur on $V$. Each node $v$ possesses a set of events $Q_v \subseteq Q$ which have occurred on it. For an event $a \in Q$, we denote the set of nodes having $a$ as $V_a$. In this paper, we use $a$ and $b$ to denote the two events for which we want to assess the structural correlation. For the sake of simplicity, we assume $G$ is undirected and unweighted. Nevertheless, the proposed approach could be extended for graphs with directed and/or weighted edges.

**Problem Statement** *Given two events $a$ and $b$ and their corresponding occurrences $V_a$ and $V_b$, to determine whether $a$ and $b$ are correlated (if correlated, positive or negative) in the graph space with respect to a vicinity level $h$.*

We formally define the notion of vicinity on a graph as follows.

DEFINITION 1 (NODE LEVEL-$h$ VICINITY). *Given graph $G = (V, E)$ and a node $u \in V$, the level-$h$ vicinity (or $h$-vicinity) of $u$ is defined as the subgraph induced by the set of nodes whose distances from $u$ are less than or equal to $h$. We use $V_u^h$ and $E_u^h$ to denote the sets of nodes and edges in $u$'s $h$-vicinity, respectively.*

DEFINITION 2 (NODE SET $h$-VICINITY). *Given a graph $G = (V, E)$ and a node set $V' \subseteq V$, the $h$-vicinity of $V'$ is defined as the subgraph induced by the set of nodes which are within distance $h$ from at least one node $u \in V'$. For event $a$, we use $V_a^h$ and $E_a^h$ to denote the sets of nodes and edges in $V_a$'s $h$-vicinity, respectively.*

Let $V_{a \cup b} = V_a \cup V_b$ denote the set of nodes having at least one of events $a$ and $b$, i.e. all event nodes. The sets of nodes and edges in the $h$-vicinity of $V_{a \cup b}$ is denoted by $V_{a \cup b}^h$ and $E_{a \cup b}^h$, respectively. To assess the structural correlation between $a$ and $b$, we employ a set of reference nodes.

DEFINITION 3 (REFERENCE NODES). *Given two events $a$ and $b$ on $G$, a node $r \in V$ is a reference node for assessing level-$h$ TESC between $a$ and $b$ iff $r \in V_{a \cup b}^h$.*

Definition 3 indicates that we treat $V_{a \cup b}^h$ as the set of all reference nodes for assessing level-$h$ TESC between $a$ and $b$. The reason will be explained in Section 3.2. We define the notion of concordance for a pair of reference nodes as follows.

DEFINITION 4 (CONCORDANCE). *Two reference nodes $r_i$ and $r_j$ for assessing level-$h$ TESC between $a$ and $b$ are said to be concordant iff both $a$'s density and $b$'s density increase (or decrease) when we move from $r_i$'s $h$-vicinity to $r_j$'s $h$-vicinity.*

Mathematically, the concordance function $c(r_i, r_j)$ is defined as

$$c(r_i, r_j) = \begin{cases} 1 & (s_a^h(r_i) - s_a^h(r_j))(s_b^h(r_i) - s_b^h(r_j)) > 0 \\ -1 & (s_a^h(r_i) - s_a^h(r_j))(s_b^h(r_i) - s_b^h(r_j)) < 0 \\ 0 & otherwise \end{cases}, \quad (1)$$

where $s_a^h(r_i)$ is the density of event $a$ in $r_i$'s $h$-vicinity:

$$s_a^h(r_i) = \frac{|V_a \cap V_{r_i}^h|}{|V_{r_i}^h|}. \quad (2)$$

$c(r_i, r_j)$ encodes concordance as 1 and discordance as -1. 0 means $r_i$ and $r_j$ are in a tie, i.e. $s_a^h(r_i) = s_a^h(r_j)$ or $s_b^h(r_i) = s_b^h(r_j)$, which means the pair indicates neither concordance nor discordance. Regarding $s_a^h(r_i)$, the reason that we use $|V_{r_i}^h|$ to normalize the occurrence number is that different nodes could have quite different sizes of $h$-vicinities. $|V_{r_i}^h|$ can be regarded as an analogue to area in spatial spaces. Normalization makes all reference nodes' $h$-vicinities have the same "area". The computation of $s_a^h(r_i)$ is simple: we do a Breadth-First Search (BFS) up to $h$ hops (hereafter to be called $h$-hop BFS) from $r_i$ to count the number of occurrences of the event. More sophisticated graph proximity measures could be used here, such as hitting time [19] and personalized PageRank [6]. However, the major issue with these sophisticated measures is the high computational cost. As will be demonstrated in experiments, our density measure is not only much more efficient but also effective.

## 3. MEASURING TESC

This section presents our TESC testing framework. First, we show the intuition behind using reference nodes to assess TESC. If events $a$ and $b$ are positively correlated on $G$, a region where $a$ appears tends to also contain occurrences of $b$, and visa versa. Furthermore, more occurrences of one event will tend to imply more occurrences of the other one. On the contrary, when $a$ and $b$ are negatively correlated, the presence of one event is likely to imply the absence of the other one. Even if they appear together, an increase of occurrences of one event is likely to imply a decrease of the other. Figure 2 shows the four typical scenarios described above. $r_1$ and $r_2$ are two reference nodes. Here let us assume $h$-vicinities (denoted by dotted circles) of $r_1$ and $r_2$ have the same number of nodes so that we can treat the number of occurrences as density. We can see in Figure 2(a) and 2(b), when $a$ and $b$ attract each other, $r_1$ and $r_2$ are concordant, implying an evidence of positive correlation. In the repulsion cases (Figure 2(c) and 2(d)), $r_1$ and $r_2$ are discordant, showing an evidence of negative correlation. Therefore, the idea is to aggregate all these evidences from all pairs of reference nodes to assess TESC.

The natural choice for computing the overall concordance among reference nodes with regard to density changes of the two events is the Kendall's $\tau$ rank correlation [15], which was also successfully applied to the spatial point pattern correlation problem [7, 23]. For clarity, let $N = |V_{a \cup b}^h|$. We have $N$ reference nodes: $r_1, r_2, \ldots, r_N$. The Kendall's $\tau$ measure is defined as an aggregation of $c(r_i, r_j)$'s

$$\tau(a, b) = \frac{\sum_{i=1}^{N-1} \sum_{j=i+1}^{N} c(r_i, r_j)}{\frac{1}{2} N(N-1)}. \quad (3)$$

$\tau(a, b)$ lies in $[-1, 1]$. A higher positive value of $\tau(a, b)$ means a stronger positive correlation, while a lower negative value

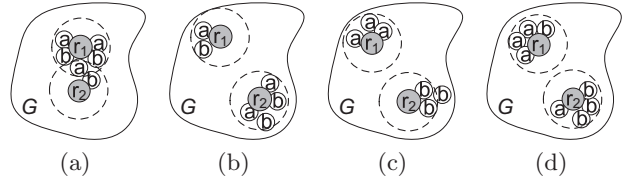

Figure 2: Four illustrative examples showing that density changes of the two events between two reference nodes show an evidence of correlation.

means a stronger negative correlation. $\tau(a, b) = 0$ means there is no correlation between $a$ and $b$, i.e. the number of evidences for positive correlation is equal to that for negative correlation.

### 3.1 The Test

If $N$ is not large, we can directly compute $\tau(a, b)$ and judge whether there is a correlation (and how strong) by $\tau(a, b)$. However, real-life graphs usually have very large sizes and so does $N$. It is often impractical to compute $\tau(a, b)$ directly. We propose to sample reference nodes and perform hypothesis testing [24] to efficiently estimate TESC. In a hypothesis test, a *null hypothesis* $H_0$ is tested against an *alternative hypothesis* $H_1$. The general process is that we compute from the sample data a statistic measure $\mathcal{X}$ which has an associated *rejection region* $C$ such that, if the measure score falls in $C$, we reject $H_0$, otherwise $H_0$ is not rejected. The *significance level* of a test, $\alpha$, is the probability that $\mathcal{X}$ falls in $C$ when $H_0$ is true. The *p-value* of a test is the probability of obtaining a value of $\mathcal{X}$ at least as extreme as the one actually observed, assuming $H_0$ is true. In our case $\mathcal{X}$ is $\tau$ and $H_0$ is: events $a$ and $b$ are independent with respect to $G$'s structure. The test methodology is as follows: first we choose uniformly a random sample of $n$ reference nodes from $V_{a \cup b}^h$; then we compute the $\tau$ score over sampled reference nodes (denoted by $t(a, b)$):

$$t(a, b) = \frac{\sum_{i=1}^{n-1} \sum_{j=i+1}^{n} c(r_{k_i}, r_{k_j})}{\frac{1}{2} n(n-1)}, \quad (4)$$

where $r_{k_1}, \ldots, r_{k_n}$ are the $n$ sampled reference nodes; finally we estimate the significance of $t(a, b)$ and reject $H_0$ if the p-value is less than a predefined significance level. We use $\mathbf{s}_a^h$ to represent the vector containing densities of $a$ measured in all $n$ sample reference nodes' $h$-vicinities where the $i$-th element is $s_a^h(r_{k_i})$. Under $H_0$, $\tau(a, b)$ is 0. Consequently, for a uniformly sampled set of reference nodes, any ranking order of $\mathbf{s}_b^h$ is equally likely for a given order of $\mathbf{s}_a^h$. It is proved that the distribution of $t(a, b)$ under the null hypothesis tends to the normal distribution with mean 0 and variance

$$\sigma^2 = \frac{2(2n+5)}{9n(n-1)}. \quad (5)$$

The idea of the proof is to show the moments of $t$'s distribution under $H_0$ converge to those of the normal distribution, and then apply the Second Limit Theorem [9]. Readers could refer to Chapter 5 of [15] for details. A good normality approximation can be obtained when $n > 30$ [15]. When $s_a^h(r_{k_i}) = s_a^h(r_{k_j})$ or $s_b^h(r_{k_i}) = s_b^h(r_{k_j})$, $c(r_{k_i}, r_{k_j})$ can be 0. This means there could be ties of reference nodes where pairs in a tie show evidences of neither concordance nor discordance. When ties are present in $\mathbf{s}_a^h$ and/or $\mathbf{s}_b^h$ (often, the case is that a set of reference nodes only have occurrences of $a$ or $b$ in their $h$-vicinities), $\sigma^2$ should be modified accordingly. Let $l/m$ be the number of ties in $\mathbf{s}_a/\mathbf{s}_b$. The variance of the numerator of Eq. (4) becomes [15]:

$$\sigma_c^2 = \frac{1}{18}[n(n-1)(2n+5) - \sum_{i=1}^{l} u_i(u_i-1)(2u_i+5)$$
$$- \sum_{i=1}^{m} v_i(v_i-1)(2v_i+5)] + \frac{1}{9n(n-1)(n-2)}$$
$$\times [\sum_{i=1}^{l} u_i(u_i-1)(u_i-2)][\sum_{i=1}^{m} v_i(v_i-1)(v_i-2)]$$
$$+ \frac{1}{2n(n-1)}[\sum_{i=1}^{l} u_i(u_i-1)][\sum_{i=1}^{m} v_i(v_i-1)], \quad (6)$$

where $u_i$ and $v_i$ are the sizes of the $i$-th ties of $\mathbf{s}_a^h$ and $\mathbf{s}_b^h$, respectively. When these sizes all equal to 1, Eq. (6) reduces to Eq. (5) multiplied by $[\frac{1}{2}n(n-1)]^2$, i.e. the variance of the numerator of Eq. (4) when no ties exist. By grouping terms involving $u_i/v_i$ together, one can verify that more (larger) ties always lead to smaller $\sigma_c^2$. $\sigma^2$ is then modified as $\sigma_c^2$ divided by $[\frac{1}{2}n(n-1)]^2$. Once the variance is obtained, we compute the significance (z-score) of the observed $t(a,b)$ by

$$z(a,b) = \frac{t(a,b) - E(t(a,b))}{\sqrt{Var(t(a,b))}} = \frac{t(a,b)}{\sigma}. \quad (7)$$

For $\tau$ we do not substitute the alternative normalization term (see Chapter 3 of [15]) for $[\frac{1}{2}N(N-1)]$ when ties are present, since it makes no difference on the significance result, i.e. simultaneously dividing $\sum_{i=1}^{n-1}\sum_{j=i+1}^{n} c(r_{k_i}, r_{k_j})$ and $\sigma_c$ by the same normalization term. $t$ is an unbiased and consistent estimator for $\tau$. In practice, we do not need to sample too many reference nodes since the variance of $t$ is upper bounded by $\frac{2}{n}(1-\tau^2)$ [15], regardless of $N$.

### 3.2 Reference Nodes

Given the occurrences of two events $a$ and $b$ on graph $G$, not all nodes in $G$ are eligible to be reference nodes for the correlation estimation between $a$ and $b$. We do not consider areas on $G$ where we cannot "see" any occurrences of $a$ or $b$. That is, we do not consider nodes whose $h$-vicinities do not contain any occurrence of $a$ or $b$. We refer to this kind of nodes as out-of-sight nodes. The reasons are: (1) we measure the correlation of presence, but not the correlation of absence. The fact that an area does not contain $a$ and $b$ currently does not mean it will never have $a$ and/or $b$ in the future; (2) if we incorporate out-of-sight nodes into our reference set, we could get unexpected high z-scores, since in that case we take the correlation of absence into account. Out-of-sight nodes introduce two 0-ties containing the same set of nodes into $\mathbf{s}_a^h$ and $\mathbf{s}_b^h$, respectively. As shown in the toy example of Figure 3, the two 0-ties contain $r_6$ through $r_9$. Adding $r_6$ through $r_9$ to the reference set can only increase the number of concordant pairs, thus increasing $\sum_{i=1}^{n-1}\sum_{j=i+1}^{n} c(r_{k_i}, r_{k_j})$. Moreover, the variance of $\sum_{i=1}^{n-1}\sum_{j=i+1}^{n} c(r_{k_i}, r_{k_j})$ under the null hypothesis is relatively reduced (Eq. (6)). These two factors tend to lead to an overestimated z-score. Therefore, given two events $a$ and $b$, we treat $V_{a\cup b}^h$ as the set of all reference nodes for assessing level-$h$ TESC between $a$ and $b$. It means we should sample reference nodes within $V_{a\cup b}^h$, otherwise we would get out-of-sight nodes. This is different from the spatial point pattern correlation problem where point patterns are assumed to be isotropic and we can easily identify and focus on regions containing points. In the next section, we study how we can do reference node sampling efficiently.

|  | $r_1$ | $r_2$ | $r_3$ | $r_4$ | $r_5$ | $r_6$ | $r_7$ | $r_8$ | $r_9$ |
|---|---|---|---|---|---|---|---|---|---|
| $\mathbf{s}_a$ | [0.0, | 0.3, | 0.1, | 0.0, | 0.4, | 0.0, | 0.0, | 0.0, | 0.0] |
| $\mathbf{s}_b$ | [0.4, | 0.6, | 0.0, | 0.7, | 0.8, | 0.0, | 0.0, | 0.0, | 0.0] |

**Figure 3:** $\mathbf{s}_a$ and $\mathbf{s}_b$ when we incorporate nodes whose $h$-vicinities do not contain any occurrence of $a$ or $b$.

## 4. REFERENCE NODE SAMPLING

In this section we present efficient algorithms for sampling reference nodes from $V_{a\cup b}^h$. We need to know which nodes are within $V_{a\cup b}^h$, but only have $V_{a\cup b}$ in hand. For continuous spaces, we can perform range search efficiently by building R-tree [3] or k-d tree [5] indices. However, for graphs it is difficult to build efficient index structures for answering range queries, e.g. querying for all nodes in one node's $h$-vicinity. Pre-computing and storing pairwise shortest distances is not practical either, since it requires $O(|V|^2)$ storage. In the following, we first propose an approach which employs BFS to retrieve all nodes in $V_{a\cup b}^h$, and then randomly chooses $n$ nodes from $V_{a\cup b}^h$. Then we present efficient sampling algorithms which avoid enumerating all nodes in $V_{a\cup b}^h$. Finally, we analyze time complexity of these algorithms.

### 4.1 Batch_BFS

The most straightforward method for obtaining a uniform sample of reference nodes is to first obtain $V_{a\cup b}^h$, and then simply sample from it. $V_{a\cup b}^h$ can be obtained by performing a $h$-hop BFS search from each node $v \in V_{a\cup b}$ and doing set unions. However, this strategy would perform poorly since the worst case time complexity is $O(|V_{a\cup b}|(|V|+|E|))$. The problem is that the $h$-vicinities of nodes in $V_{a\cup b}$ could have many overlaps. Therefore, we adopt a variant of $h$-hop BFS search which starts with all nodes in $V_{a\cup b}$ as source nodes. For clarity, we show the algorithm Batch_BFS in Algorithm 1. It is similar to the $h$-hop BFS algorithm for one source node, except that the queue $Queue$ is initialized with a set of nodes. The correctness of Batch_BFS can be easily verified by imagining that we do a $(h+1)$-hop BFS from a virtual node which is connected to all nodes in $V_{a\cup b}$. By means of Batch_BFS, the worst case time complexity is reduced from $O(|V_{a\cup b}|(|V|+|E|))$ to $O(|V|+|E|)$, which means for each node in the graph we do adjacency list examination at most once. As we will show in experiments, though simple, Batch_BFS is a competitive method for reference node selection.

### 4.2 Importance Sampling

Though Batch_BFS algorithm is efficient in that its worst case time cost is *linear* in the number of nodes plus the number of edges in the graph, it still enumerates all $N$ reference nodes. In practice, the sample size $n$ is usually much smaller

**Algorithm 1:** Batch_BFS

**Input**: $Adj\_lists$: Adjacency lists for all nodes in $G$, $V_{a \cup b}$: the set of all event nodes, $h$: # of hops
**Output**: $V_{out}$: all nodes in $h$-vicinity of $V_{a \cup b}$
**begin**
    Initialize $V_{out} = \emptyset$.
    Initialize queue $Queue$ with all $v \in V_{a \cup b}$ and set $v.depth = 0$.
    **while** $Queue$ is not empty **do**
        $v = Dequeue(Queue)$
        **foreach** $u$ in $Adj\_lists(v)$ **do**
            **if** $u \notin V_{out}$ and $u \notin Queue$ **then**
                $u.depth = v.depth + 1$
                **if** $u.depth \geqslant h$ **then**
                      $V_{out} = V_{out} \cup \{u\}$
                **else**
                      $Enqueue(Queue, u)$
                **end**
            **end**
        **end**
        $V_{out} = V_{out} \cup \{v\}$
    **end**
**end**

---

**Procedure** RejectSamp($V_{a \cup b}$)

1. Select a node $v \in V_{a \cup b}$ with probability $|V_v^h|/N_{sum}$.
2. Sample a node $u$ from $V_v^h$ uniformly.
3. Get the number of event nodes in $u$'s $h$-vicinity: $c = |V_u^h \cap V_{a \cup b}|$.
4. Flip a coin with success probability $\frac{1}{c}$. Accept $u$ if we succeed, otherwise a failure occurs.

---

than $N$ and can be treated as a constant since we can fix $n$ for testing different pairs of events. Hence, the question is, can we develop reference node selection algorithms which have time costs depending on $n$, rather than $N$?

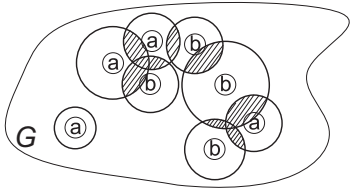

**Figure 4:** $h$-vicinities of event nodes.

The idea is that we directly sample reference nodes without firstly enumerating the whole set of reference nodes. It is challenging since we want to sample from the uniform probability distribution over $V_{a \cup b}^h$, but only have $V_{a \cup b}$ in hand. The basic operation is randomly picking an event node in $V_{a \cup b}$ and peeking at its $h$-vicinity. It is not easy to achieve uniform sampling. On one hand, the $h$-vicinities of event nodes could have many overlapped regions, as illustrated by Figure 4. Circles represent $h$-vicinities of the corresponding nodes and shadowed regions are overlaps. Nodes in overlapped regions are easier to be selected if we sample nodes uniformly from a random event node's $h$-vicinity. On the other hand, different nodes have $h$-vicinities with different node set sizes, i.e. $|V_v^h|$, conveyed by circle sizes in Figure 4. If we pick event nodes uniformly at random, nodes in small circles tend to have higher probabilities to be chosen.

We can use rejection sampling [10] to achieve uniform sampling in $V_{a \cup b}^h$, if we know $|V_v^h|$ for each $v \in V_{a \cup b}$. Let $N_{sum} = \sum_{v \in V_{a \cup b}} |V_v^h|$ be the sum of node set sizes of all event nodes' $h$-vicinities. It is easy to verify $N_{sum} \geq N$ due to overlaps. The sampling procedure is shown in Procedure RejectSamp. Proposition 1 shows that RejectSamp generates samples from the uniform probability distribution over $V_{a \cup b}^h$. $|V_v^h|$'s ($h = 1, \ldots, h_m$) can be pre-computed offline by doing a $h_m$-hop BFS from each node in the graph. The space cost is only $O(|V|)$ for each vicinity level and once we obtain the index, it can be efficiently updated as the graph changes. The time cost depends on $|V|$ and the average size of node $h_m$-vicinities, i.e. average $|V_v^{h_m}| + |E_v^{h_m}|$. Fortunately, we do not need to consider too high values of $h$ since (1) correlations of too broad scales usually do not convey useful information and (2) in real networks like social networks, increasing $h$ would quickly let a node's $h$-vicinity cover a large fraction of the network due to the "small world" phenomenon of real-life networks [2]. Therefore, we focus on relatively small $h$ values, such as $h = 1, 2, 3$.

PROPOSITION 1. RejectSamp *generates each node in $V_{a \cup b}^h$ with equal probability.*

PROOF. Consider an arbitrary node $u \in V_{a \cup b}^h$. In step 2 of RejectSamp, $u$ has a chance to be sampled if a node $v \in V_u^h \cap V_{a \cup b}$ is selected in step 1. Thus, the probability that $u$ is generated after step 2 is $\sum_{v \in V_u^h \cap V_{a \cup b}} \frac{|V_v^h|}{N_{sum}} \times \frac{1}{|V_v^h|} = \frac{|V_u^h \cap V_{a \cup b}|}{N_{sum}}$. This is a non-uniform probability distribution over $V_{a \cup b}^h$. Then by the discount in step 4, $u$ is finally generated with probability $\frac{1}{N_{sum}}$, which is independent of $u$. □

Each run of RejectSamp incurs a cost of two $h$-hop BFS searches (step 2 and 3). Simply repeating RejectSamp until $n$ reference nodes are obtained will generate a uniform sample of reference nodes. However, each run of RejectSamp could fail. The success probability of a run of RejectSamp is $p_{\text{succ}} = N/N_{sum}$, which can be easily derived by aggregating success probabilities of all nodes in $V_{a \cup b}^h$. When there is no overlap among event nodes' $h$-vicinities, $p_{\text{succ}} = 1$ since $N_{sum} = N$. The expected time cost in terms of $h$-hop BFS is $2n/p_{\text{succ}}$. It means the heavier the overlap among different event nodes' $h$-vicinities is, the higher the cost is. Considering the "small world" property of real-life networks [2], it would be easy to get a heavy overlap as $V_{a \cup b}$ and $h$ grow. Preliminary experiments confirm RejectSamp is inefficient.

We propose a weighting technique to address the above problem. The idea is similar to importance sampling [13]. In particular, we use the same sampling scheme with RejectSamp except that we do not reject any sampled nodes. This leads to samples generated from the nonuniform distribution $\mathbb{P} = \{p(v)\}_{v \in V_{a \cup b}^h}$ where $p(v) = |V_v^h \cap V_{a \cup b}|/N_{sum}$. Notice that $t(a, b)$ is intrinsically an estimator of the real correlation score $\tau(a, b)$. The idea is, if we can derive a proper estimator for $\tau(a, b)$ based on samples from $\mathbb{P}$, we could use it as a surrogate to $t(a, b)$. Let $\mathcal{S} = \{(r_1, w_1), \ldots, (r_n, w_n)\}$ be a set consisting of $n$ distinct reference nodes sampled from $\mathbb{P}$, where $w_i$ is the number of times $r_i$ is generated in the sampling process. We denote the sample size of $\mathcal{S}$ as $n' = \sum_{i=1}^{n} w_i$. We define a new estimator for $\tau(a, b)$ based on $\mathcal{S}$

$$\tilde{t}(a,b) = \frac{\sum_{i=1}^{n-1} \sum_{j=i+1}^{n} c(r_i, r_j) \frac{w_i w_j}{p(r_i)p(r_j)}}{\sum_{i=1}^{n-1} \sum_{j=i+1}^{n} \frac{w_i w_j}{p(r_i)p(r_j)}}. \quad (8)$$

**Algorithm 2:** Importance sampling

**Input**: $V_{a\cup b}$: the set of all event nodes, $|V_v^h|$: $h$-vicinity node set sizes for all $v \in V_{a\cup b}$, $h$: # of hops
**Output**: $\mathcal{S}$: a set of $n$ sampled reference nodes, $W$: the set of weights (frequencies) for each $r \in \mathcal{S}$

1 **begin**
2     Initialize $\mathcal{S} = \emptyset$.
3     **while** $|\mathcal{S}| < n$ **do**
4        Randomly select a node $v \in V_{a\cup b}$ with probability $|V_v^h|/N_{sum}$.
5        Do a $h$-hop BFS search from $v$ to get $V_v^h$ and sample a node $r$ from $V_v^h$ uniformly.
         **if** $r \in \mathcal{S}$ **then**
            $W(r) = W(r) + 1$
         **else**
            $\mathcal{S} = \mathcal{S} \cup \{r\}$
            $W(r) = 1$
         **end**
    **end**
**end**

**Algorithm 3:** Whole graph sampling

**Input**: $V_{a\cup b}$: the set of all event nodes, $V$: all nodes in graph $G$
**Output**: $\mathcal{S}$: a set of $n$ sampled reference nodes

1 **begin**
2     $\mathcal{S} = \emptyset$
3     **while** $|\mathcal{S}| < n$ **do**
4        Randomly pick a node $v \in V$
5        Do a $h$-hop BFS search from $v$ to get $V_v^h$
         **if** $V_v^h \cap V_{a\cup b} \neq \emptyset$ **then**
            $\mathcal{S} = \mathcal{S} \cup \{v\}$
         **end**
         $V = V - \{v\}$
    **end**
**end**

This estimator is a consistent estimator of $\tau(a,b)$, which is proved in Theorem 1.

THEOREM 1. *$\tilde{t}(a,b)$ is a consistent estimator of $\tau(a,b)$.*

PROOF. To prove $\tilde{t}(a,b)$ is a consistent estimator for $\tau(a,b)$, we need to show that $\tilde{t}(a,b) \xrightarrow{P} \tau(a,b)$, i.e. $\tilde{t}(a,b)$ converges to $\tau(a,b)$ in probability as the sample size $n' \to \infty$. For each $r_i$, we define a Bernoulli random variable $X_{r_i}$ which is 1 if a run of sampling from $\mathbb{P}$ outputs node $r_i$, and 0 otherwise. $\frac{w_i}{n'}$ is the sample mean for $X_{r_i}$. By the Law of Large Numbers, as $n' \to \infty$, $\frac{w_i}{n'}$ converges in probability to the expectation $E(X_{r_i}) = p(r_i)$. Moreover, all nodes in $V_{a\cup b}^h$ will be added into $\mathcal{S}$ when $n' \to \infty$, which means $n = N$. Therefore, as $n' \to \infty$, we can obtain:

$$\tilde{t}(a,b) = \frac{\frac{1}{n'^2}\sum_{i=1}^{N-1}\sum_{j=i+1}^{N} c(r_i, r_j) \frac{w_i w_j}{p(r_i)p(r_j)}}{\frac{1}{n'^2}\sum_{i=1}^{N-1}\sum_{j=i+1}^{N} \frac{w_i w_j}{p(r_i)p(r_j)}}$$
$$= \frac{\sum_{i=1}^{N-1}\sum_{j=i+1}^{N} c(r_i, r_j) \frac{p(r_i)p(r_j)}{p(r_i)p(r_j)}}{\sum_{i=1}^{N-1}\sum_{j=i+1}^{N} \frac{p(r_i)p(r_j)}{p(r_i)p(r_j)}} = \tau(a,b),$$

which completes the proof. □

It is easy to verify that $\tilde{t}(a,b)$ is a biased estimator by considering a toy problem and enumerating all possible outputs of a sample of size $n'$ (together with their probabilities) to compute $E(\tilde{t}(a,b))$. However, unbiasedness used to receive much attention but nowadays is considered less important [24]. We will empirically demonstrate that $\tilde{t}(a,b)$ can achieve acceptable performance in experiments. For clarity, we show the Importance sampling algorithm in Algorithm 2. In each iteration of the sampling loop, the major cost is one $h$-hop BFS search (line 5). The number of iterations $n'$, though $\geq n$, is typically $\approx n$ in practice. This is because when $N$ is large, the probability of selecting the same node in different iterations is very low. Thus, the major cost of Importance sampling could be regarded as depending on $n$. Once $\mathcal{S}$ and $W$ are obtained, we can then compute $\tilde{t}(a,b)$ as a surrogate for $t(a,b)$ and assess the significance accordingly.

**Improving Importance Sampling** Although the time cost of Importance sampling depends on $n$ rather than $N$, in practice $n$ $h$-hop BFS searches could still be slower than one Batch_BFS search as $h$ increases. This is because the overlap among different event nodes' $h$-vicinities tends to become heavier as $h$ increases. We can alleviate this issue by sampling reference nodes in a *batch* fashion. That is, when $V_v^h$ is obtained for a sampled $v \in V_{a\cup b}$ (line 5 of Algorithm 2), we sample more than one reference nodes from $V_v^h$. In this way, the ratios between different reference nodes' probabilities of being chosen do not change. However, this also introduces dependence into $\mathcal{S}$. Sampling too many nodes from one $V_v^h$ would degrade performance since the number of event nodes peeked at decreases and consequently we are more likely to be trapped in local correlations. This is a tradeoff between efficiency and accuracy. We will test this approximation idea in experiments.

### 4.3 Global Sampling in Whole Graph

When $|V_{a\cup b}|$ and $h$ increase, the chance that a random node selected from the whole graph is in $V_{a\cup b}^h$ also increases. In this situation, we can simply sample nodes uniformly in the whole graph and the obtained nodes which are within $V_{a\cup b}^h$ can be regarded as a uniform sample from $V_{a\cup b}^h$. We use an iterative process to harvest reference nodes: (1) firstly a node is chosen uniformly from the whole graph; (2) test whether the selected node is within $V_{a\cup b}^h$; (3) if it is in $V_{a\cup b}^h$, keep it. (4) another node is selected uniformly from the remaining nodes and we go to step 2. This process continues until $n$ reference nodes are collected. For completeness, the Whole graph sampling algorithm is shown in Algorithm 3. The major cost is incurred by one $h$-hop BFS search in each iteration (line 5), where the purpose is to examine whether $v$ is an eligible reference node.

### 4.4 Complexity Analysis

The major space cost is $O(|E|)$, for storing the graph as adjacency lists. Regarding time complexity, we have mainly three phases: reference node sampling, event density computation (Eq. (2)) and measure computation (z-score, Eq. (7)). Let $c_B$ be the average cost of one $h$-hop BFS search on graph $G$, which is linear in the average size of node $h$-vicinities, i.e. average $|V_v^h| + |E_v^h|$. Let $n$ be the number of sample reference nodes. The event density computation for a reference node has time complexity $O(c_B)$. The cost of z-score computation is $O(n^2)$. Fortunately, we do not need to select too many reference nodes, as discussed in Section 3.1. We will demonstrate the efficiency of the above two phases in experiments.

For reference node sampling, we have three methods. The time complexity of Batch_BFS is $O(|V_{a\cup b}^h| + |E_{a\cup b}^h|)$ where $|V_{a\cup b}^h| = N$. The cost of Importance sampling is $O(nc_B)$. For Whole graph sampling, the time cost is $O(n_f c_B)$, where $n_f$ is the number of nodes examined which are not in $V_{a\cup b}^h$. The cost incurred by examined nodes which are in $V_{a\cup b}^h$ is counted in the event density computation phase. $n_f$ is a random variable. Treating Whole graph sampling as sampling with replacement, the probability of selecting a node in $V_{a\cup b}^h$ in each iteration is $N/|V|$. The expected total number of iterations is $n|V|/N$ and therefore $E(n_f) = n|V|/N - n$. When $N$ is small, Batch_BFS can be used. For large $N$, Importance sampling and Whole graph sampling are better candidates. We will empirically analyze their efficiency in the experiments.

## 5. EXPERIMENTS

This section presents the experimental results of applying our proposed TESC testing framework on several real world graph datasets. Firstly, we verify the efficacy of the proposed TESC testing framework by event simulation on the DBLP graph. Then we examine the efficiency and scalability of the framework with a Twitter network. The third part of experiments concentrates on analyzing highly correlated real event pairs discovered by our measure in real graph datasets. All experiments are run on a PC with Intel Core i7 CPU and 12GB memory.

### 5.1 Graph Datasets

We use three datasets to evaluate our TESC testing framework: DBLP, Intrusion and Twitter.

**DBLP** The DBLP dataset was downloaded on Oct. 16th, 2010 (http://www.informatik.uni-trier.de/∼ley/db). Its paper records were parsed to obtain the co-author social network. Keywords in paper titles are treated as events associated with nodes (authors) on the graph. The DBLP graph contains 964,677 nodes and 3,547,014 edges. Totally, it has around 0.19 million keywords.

**Intrusion** The Intrusion dataset was derived from the log data of intrusion alerts in a computer network. It has 200,858 nodes and 703,020 edges. There are 545 different types of alerts which are treated as events in this network.

**Twitter** The Twitter dataset has 20 million nodes and 0.16 billion edges, which is a bidirectional subgraph of the whole twitter network (http://twitter.com). We do not have events for this dataset. It is used to test the scalability of the proposed TESC testing framework.

### 5.2 Event Simulation

A suitable method for evaluating the efficacy of our approach is to simulate correlated events on graphs and see if we can correctly detect correlations. Specifically, we adopt similar methodologies as those used in the analogous point pattern problem [7] to generate pairs of events with positive and negative correlations on graphs. DBLP network is used as the test bed. We investigate correlations with respect to different vicinity levels $h = 1, 2, 3$. Positively correlated event pairs are generated in a linked pair fashion: we randomly select 5000 nodes from the graph as event $a$ and each node $v \in V_a$ has an associated event $b$ node whose distance to $v$ is described by a Gaussian distribution with mean zero and variance equal to $h$ (distances go beyond $h$ are set to $h$). When the distance is decided, we randomly pick a node at that distance from $v$ as the associated event $b$ node. This represents strong positive correlations since wherever we observe an event $a$, there is always a nearby event $b$. For negative correlation, again we first generate 5000 event $a$ nodes randomly, after which we employ Batch_BFS to retrieve the nodes in the $h$-vicinity of $V_a$, i.e. $V_a^h$. Then we randomly color 5000 nodes in $V \setminus V_a^h$ as having event $b$. In this way, every node of $b$ is kept at least $h+1$ hops away from all nodes of $a$ and the two events exhibit a strong negative correlation. For each vicinity level, we generate 100 positive event pairs and 100 negative event pairs from the above simulation processes, respectively. We use *recall* as the evaluation metric which is defined as the number of correctly detected event pairs divided by the total number of event pairs (100). We report results obtained from one-tailed tests with significance level $\alpha = 0.05$. In our experiments, we empirically set the sample size of reference nodes $n = 900$.

#### 5.2.1 Performance Comparison

We investigate the performance of three reference node sampling algorithms, namely, Batch_BFS, Importance sampling and Whole graph sampling, under different vicinity levels and different noise levels. Noises are introduced as follows. Regarding positive correlation, we introduce a sequence of independent Bernoulli trails, one for each linked pair of event nodes, in which with probability $p$ the pair is broken and the node of $b$ is relocated outside $V_a^h$. For negative correlation, given an event pair each node in $V_b$ has probability $p$ to be relocated and attached with one node in $V_a$. The probability $p$ controls to what extent noises are introduced and can be regarded as noise level.

We show the experimental results in Figure 5 and 6, for positive correlation and negative correlation, respectively. As can be seen, overall the performance curves start from 100% and fall off as the noise level increases. This indicates that the proposed statistical testing approach is efficacious for measuring TESC. Among the three reference node sampling algorithms, Batch_BFS achieves relatively better performance. Importance sampling, though not as good as Batch_BFS, can also achieve acceptable recall, especially for $h = 1, 2$. We shall show in Section 5.3 that Importance sampling is more efficient than Batch_BFS in many cases. Whole graph sampling also shows good recall in most cases, as expected. However, its running time can vary drastically and therefore it can only be applied in limited scenarios. An interesting phenomenon is that positive correlations for higher vicinity levels (e.g. 3) are harder to break than those for lower levels, while for negative correlations it is the reverse: lower level ones are harder to break. Note that the noise level ranges in subfigures of Figure 5 and 6 are not the same. This is intuitive. Consider the size of $V_a^h$. When $h$ increases, $|V_a^h|$ usually increases exponentially. For example, among our synthetic events in DBLP graph, the typical size of $V_a^1$ is 60k while that of $V_a^3$ is 700k (7/10 of the whole graph), for $|V_a| = 5000$. Hence, it is much harder for event $b$ to "escape" event $a$ for higher vicinity levels. On the contrary, for $h = 1$ it is easier to find a node whose 1-vicinity dose not even overlap with $V_a^1$. Hence, low vicinity level positive correlations and high vicinity level negative correlations are hard to maintain and consequently more inter-esting that those in other cases. In the following experiment on real events, we will focus on these interesting cases.

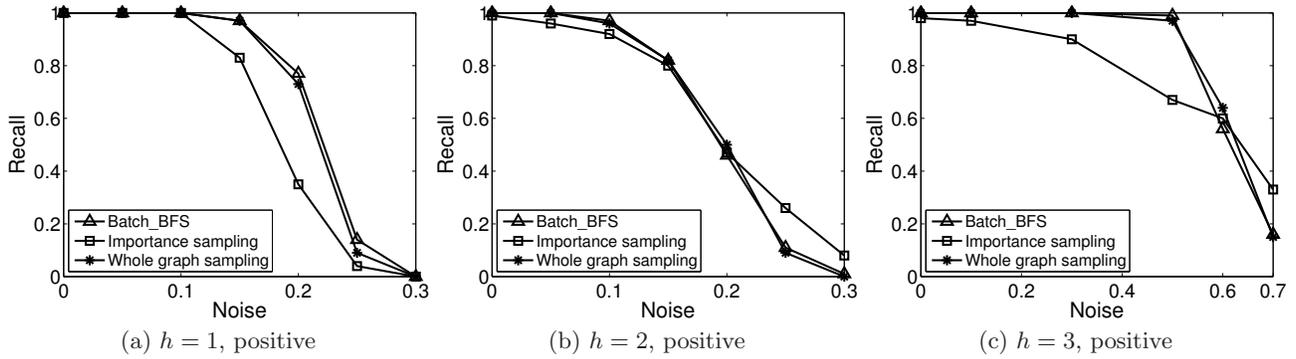

(a) $h = 1$, positive  (b) $h = 2$, positive  (c) $h = 3$, positive

Figure 5: Performance of three reference node sampling algorithms on simulated positively correlated event pairs. Results for various noise levels are reported under different vicinity levels.

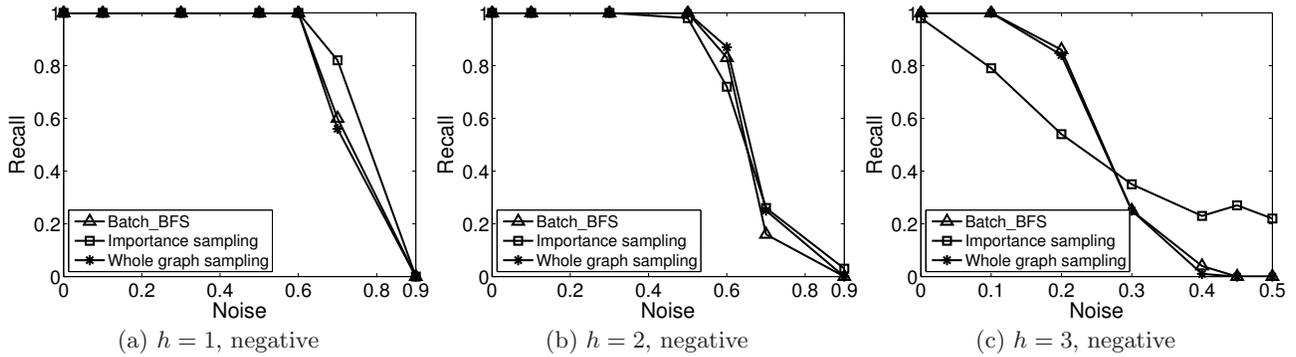

(a) $h = 1$, negative  (b) $h = 2$, negative  (c) $h = 3$, negative

Figure 6: Performance of three reference node sampling algorithms on simulated negatively correlated event pairs. Results for various noise levels are reported under different vicinity levels.

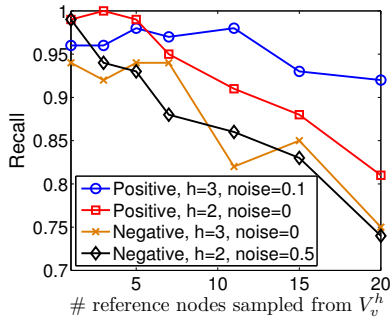

Figure 7: Performance of sampling different number of reference nodes from each $V_v^h$ for Importance sampling.

### 5.2.2 Batch Importance Sampling

In Importance Sampling, when $V_v^h$ is obtained for a sampled $v \in V_{a \cup b}$ (line 5 of Algorithm 2), we could sample more than one node from $V_v^h$ as reference nodes, in order to reduce the cost. However, sampling too many nodes from one $V_v^h$ would degrade performance since the number of event nodes peeked at decreases and consequently we are more likely to be trapped in local correlations. Here we present an empirical evaluation of this idea for $h = 2, 3$. We show results for four synthetic event pair sets in Figure 7. Two of those sets contain noises since in the corresponding cases the correlation is hard to break, which means in those cases it is easy to detect correlations. We can see that the results are as expected. The performance curves for $h = 3$ can keep high for a longer range of the number of reference nodes sampled from each $V_v^h$, compared to $h = 2$. This is because 3-vicinities are usually much larger than 2-vicinities and 3-vicinities of event nodes tend to have more overlapped regions. Therefore, sampling a batch of reference nodes from 3-vicinities is less likely to be trapped in local correlations than from 2-vicinities. The results also indicate that we can sample a small number of reference nodes from each $V_v^h$ for Importance sampling, without severely affecting its performance. In the following efficiency experiments, we set this number to 3 and 6 for $h = 2$ and $h = 3$ respectively.

### 5.2.3 Impact of Graph Density

We change the graph density to see the impact on correlation results. Specifically, we alter the DBLP graph by randomly adding/removing edges and run Batch_BFS for the six event pair sets (without noises) generated in Section 5.2. Figure 8 shows the results. We can see when removing edges, the recall of positive pairs decreases, while adding edges leads to recall decline of negative pairs. In the remaining cases (e.g. negative pairs vs. edge removal) the recall remains at 1. This is because removing edges tends to increase distances among nodes while adding edges makes nodes near one another. Figure 8(a) shows that 1-hop posi-

1407

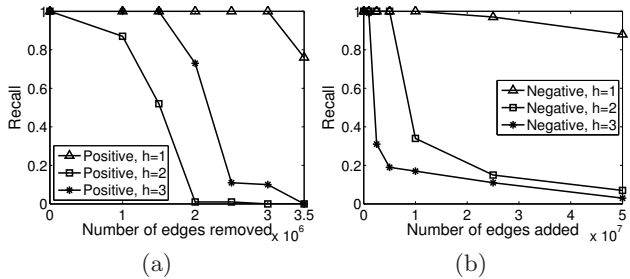

**Figure 8: Impact of randomly removing or adding edges on the correlation results.**

tive correlations are less influenced by edge removal, which is different from the observation in Section 5.2.1, i.e. 1-hop positive correlations are easier to break. The reason is that in our correlation simulation model 1-hop positive event pairs tend to have more nodes with both events, due to the Gaussian distributed distances between event $b$ nodes and corresponding event $a$ nodes. Nodes with both events reflect transaction correlation which is not influenced by edge removal. However, TESC does not just measure transaction correlations. We will show in Section 5.4 there are real event pairs which exhibit high positive TESC but are independent or even negatively correlated by transaction correlation.

### 5.3 Efficiency and Scalability

We test efficiency and scalability of our TESC testing framework on Twitter graph. First we investigate the running time of different reference node sampling algorithms with respect to the number of event nodes, i.e. the size of $V_{a \cup b}$. In particular, we randomly pick nodes from the Twitter graph to form $V_{a \cup b}$ with sizes ranging from 1000 to 500000. Then each algorithm is run to generate sample reference nodes for these $V_{a \cup b}$'s in order to record its running time. Results are averaged over 50 test instances for each size of $V_{a \cup b}$. Figure 9 shows the results for the three vicinity levels. To keep the figures clear, we do not show the running time of Whole graph sampling for some cases since its running time goes beyond 10 seconds. We can see that for different vicinity levels the situations are quite different. Generally speaking, the running time of BFS increases significantly as $V_{a \cup b}$ grows, while that of Importance sampling hardly increases. This is consistent with our analysis in Section 4.4. The running time of Importance sampling increases a little in that the algorithm tends to choose event nodes with large $V_v^h$ to peek in the sampling loop. By chance, there would be more and more event nodes with large sizes of $V_v^h$ as $V_{a \cup b}$ grows. We can see Importance sampling is definitely more efficient than Batch_BFS when $h = 1$. For $h = 2$ and 3, when the size of $V_{a \cup b}$ is small, we can use Batch_BFS; for large sizes of $V_{a \cup b}$, Importance sampling is a better choice. Whole graph sampling is recommended only for $h = 3$ and for large sizes of $V_{a \cup b}$ (above 200k in the case of Twitter graph). To conclude, the results indicate our reference sampling algorithms are efficient and scalable, i.e., we can process $V_{a \cup b}$ with 500K nodes on a graph with 20M nodes in 1.5s.

Besides reference node sampling, the TESC testing framework also needs to do one $h$-hop BFS search for each sample reference node to compute event densities and then calculate $z(a, b)$. Figure 10 shows that these two operations are also efficient and scalable. Figure 10(a) indicates that on a graph with 20 million nodes, one 3-hop BFS search only needs 5.2ms, which is much faster than the state-of-art hitting time approximation algorithm (170ms for 10 million nodes) [11]. Efficiency is the major reason that we choose this simple density measure, rather than more complicated proximity measures such as hitting time. On the other hand, although the measure computation has time complexity $O(n^2)$, we do not need to select too many reference nodes since the variance of $t(a, b)$ is upper bounded by $\frac{2}{n}(1 - \tau(a, b)^2)$ [15], regardless of $N$. Figure 10(b) shows we can compute $z(a, b)$ in 4ms for 1000 reference nodes.

**Table 1: Five keyword pairs exhibiting high 1-hop positive correlation (DBLP). All scores are z-scores.**

| # | Pair | TESC | | | TC |
|---|---|---|---|---|---|
| | | $h=1$ | $h=2$ | $h=3$ | |
| 1 | Texture vs. Image | 6.22 | 19.85 | 30.58 | 172.7 |
| 2 | Wireless vs. Sensor | 5.99 | 23.09 | 32.12 | 463.7 |
| 3 | Multicast vs. Network | 4.21 | 18.37 | 26.66 | 123.2 |
| 4 | Wireless vs. Network | 2.06 | 17.41 | 27.90 | 198.2 |
| 5 | Semantic vs. RDF | 1.72 | 16.02 | 24.94 | 120.3 |

**Table 2: Five keyword pairs exhibiting high 3-hop negative correlation (DBLP). All scores are z-scores.**

| # | Pair | TESC | | | TC |
|---|---|---|---|---|---|
| | | $h=1$ | $h=2$ | $h=3$ | |
| 1 | Texture vs. Java | -23.63 | -9.41 | -6.40 | 4.33 |
| 2 | GPU vs. RDF | -24.47 | -14.64 | -6.31 | 1.24 |
| 3 | SQL vs. Calibration | -21.29 | -12.70 | -5.45 | -0.62 |
| 4 | Hardware vs. Ontology | -22.31 | -8.85 | -5.01 | 3.38 |
| 5 | Transaction vs. Camera | -22.20 | -7.91 | -4.26 | 4.85 |

### 5.4 Real Events

We provide case studies of applying our TESC testing framework on real events occurring in real graphs. We use Batch_BFS for reference node selection. As aforementioned in Section 5.2.1, low level positive correlations and high level negative correlations are of interests. Hence, we report typical highly correlated event pairs we found in DBLP and Intrusion datasets in terms of 1-hop positive TESC and 3-hop negative TESC respectively. We report z-scores as the significance scores of the correlated event pairs. To give a notion of the correspondence between z-scores and p-values, a z-score $> 2.33$ or $< -2.33$ indicates the corresponding p-value $< 0.01$ for one-tailed significance testing. Before presenting the results, we would like to emphasize that our correlation findings are for specific networks and our measure detects exhibition of correlation, but not its cause.

Tables 1 and 2 show the results for DBLP. For comparison, we also show correlation scores measured by treating nodes as isolated transactions. We use Kendall's $\tau_b$ [1] to estimate the Transaction Correlation (TC) since $\tau_b$ can capture both positive and negative correlations. All scores in the Tables are z-scores. We can see that highly positively correlated keywords are semantically related and reflect hot research areas in different communities of computer science,

1408

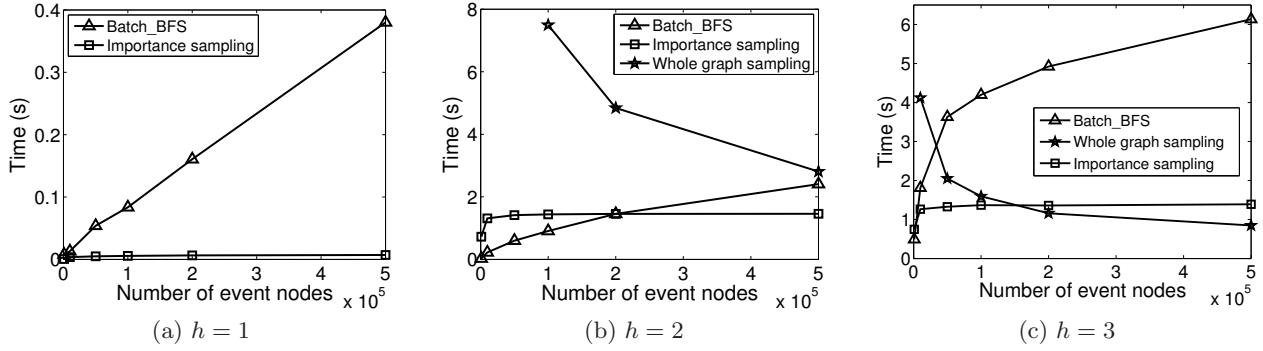

Figure 9: Running time of reference node sampling algorithms with increasing number of event nodes.

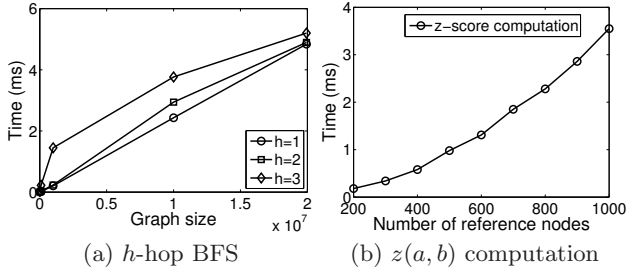

Figure 10: Running time of one $h$-hop BFS search and $z(a,b)$ computation.

Table 3: Five alert pairs exhibiting high 1-hop positive correlation (Intrusion). All scores are z-scores.

| # | Pair | TESC ($h=1$) | TC |
|---|------|---------------|-----|
| 1 | Ping_Sweep vs. SMB_Service_Sweep | 13.64 | 1.91 |
| 2 | Ping_Flood vs. ICMP_Flood | 12.53 | 5.87 |
| 3 | Email_Command_Overflow vs. Email_Pipe | 12.15 | -0.04 |
| 4 | HTML_Hostname_Overflow vs. HTML_NullChar_Evasion | 9.08 | 0.59 |
| 5 | Email_Error vs. Email_Pipe | 4.34 | -3.52 |

while negatively correlated ones represent topics which are far away from each other. In DBLP, keyword pairs having positive TESC tend to also have positive TC. However, for the negative case the results are not consistent. We can see in Table 2 many pairs have positive TC. It means although some authors have used both two keywords, they are far away in the graph space, reflecting the fact that they represent quite different topics pursued by different communities in the co-author social network.

Results for the Intrusion dataset are presented in Tables 3 and 4. Since the Intrusion graph contains several nodes with very high degrees (around 50k), its diameter is much lower than that of DBLP. In the Intrusion graph, 2-vicinity of a node tends to cover a large number of nodes. Therefore, for negative TESC we focus on $h = 2$. As shown in Table 3, positively correlated alerts reflect high-level intrusion activities. The first pair reflects pre-attack probes. The second one is related to ICMP DOS Attack. The third and fifth pairs indicate that the attacker is trying to gain root access of those hosts by vulnerabilities in email softwares and services. The fourth one is related to Internet Explorer's vulnerabilities. Notice that the third pair is nearly independent and the fifth pair is negatively correlated under TC. The reason could be that some attacking techniques consume bandwidth and there is a tradeoff between the number of hosts attacked and the number of techniques applied to one host. Attackers might choose to maximize coverage by alternating related intrusion techniques for hosts in a subnet, in order to increase the chance of success. Although these alerts represent related techniques, they do not exhibit positive TC. TESC can detect such positive structural correlations.

Table 4: Five alert pairs exhibiting high 2-hop negative correlation (Intrusion). All scores are z-scores.

| # | Pair | TESC ($h=2$) | TC |
|---|------|---------------|-----|
| 1 | Audit_TFTP_Get_Filename vs. LDAP_Auth_Failed | -31.30 | -0.81 |
| 2 | LDAP_Auth_Failed vs. TFTP_Put | -31.12 | -0.81 |
| 3 | DPS_Magic_Number_DoS vs. HTTP_Auth_TooLong | -30.96 | -0.18 |
| 4 | LDAP_BER_Sequence_Dos vs. TFTP_Put | -30.30 | -1.57 |
| 5 | Email_Executable_Extension vs. UDP_Service_Sweep | -26.93 | -0.97 |

On the other hand, highly negatively correlated alerts are those related to different attacking approaches, or in connection with different platforms. For example, in the first pair of Table 4 LDAP_Auth_Failed is related to brute-force password guessing, while Audit_TFTP_Get_Filename is related to TFTP Attack which allows remote users to write files to the target system without any authentication; in the third pair, DPS_Magic_Number_DoS is exclusive for Microsoft Dynamics GP software while HTTP_Auth_TooLong is for Netscape Enterprise Server software. These pairs also exhibit moderate negative TC.

We also compare our results with those produced by the proximity pattern mining problem [16] for the positive case. Specifically, we set $minsup = 10/|V|$ for the pFP algorithm and $\alpha = 1$, $\epsilon = 0.12$ [16]. Then we run the proximity pattern mining method on the Intrusion dataset. From the results, we find that most highly positively correlated pairs detected by **TESC** are also reported as proximity patterns, or subsets of proximity patterns. However, some rare event pairs detected by **TESC** are not discovered by the proximity pattern mining method. Table 5 shows two such examples. Digits in parentheses are event sizes. The reason is that proximity pattern mining is intrinsically a frequent pattern mining problem [16]. It requires events to occur not only closely but also frequently closely on the graph. In **TESC** there is no such requirement and we could detect positively correlated rare event pairs.

Table 5: Two rare alert pairs with positive 1-hop **TESC** which are not discovered by proximity pattern mining.

| Pair (count) | z-score/p-value |
|---|---|
| HTTP_IE_Script_HRAlign_Overflow (16) vs. HTTP_DotDotDot (29) | 3.30/0.0005 |
| HTTP_ISA_Rules_Engine_Bypass (81) vs. HTTP_Script_Bypass (12) | 2.52/0.0059 |

## 6. DISCUSSIONS

A straightforward measure for **TESC** could be to calculate the average distance between nodes of the two events. Measures of this kind try to capture the "distance" between the two events directly. However, for these direct measures it is difficult to estimate their distributions in the null hypothesis (i.e. no correlation). An empirical approach is to use randomization: perturbing events $a$ and $b$ independently in the graph with the observed sizes and internal structures, and calculating the empirical distribution of the measure. Unfortunately, it is hard to preserve each event's internal structure, thus making randomization not effective. Our approach avoids randomization by indirectly measuring the rank correlation between two events' densities in local neighborhoods of sampled reference nodes. Significance can be estimated by $\tau$'s nice property of being asymptotically normal under the null hypothesis. Our approach provides a systematic way to compute formal and rigorous statistical significance, rather than empirical one.

Another simple idea is that we first map nodes in a graph to a Euclidean space by preserving the structural properties and then apply existing techniques for spatial data. Nevertheless, (1) techniques for spatial data are not scalable; (2) mapping introduces approximation errors. For example, researchers tried to approximate network distances using a coordinate system [21, 26]. According to the recent work [26], one distance estimation costs $0.2\mu s$. Let us take the most recent method for spatial data [23] as an example. It requires estimating the distances between each reference point and all event points. Consequently, for 500K event points and 900 reference points, the total time cost is 90s! Although we could build k-d tree indices [5] to improve efficiency, k-d tree only works well for low dimensional spaces. Reducing the dimensionality leads to a higher distance estimation error [26], indicating a tradeoff between accuracy and efficiency. Our method avoids these annoying issues and provides a scalable solution over the exact structure.

How to choose the sample size of reference nodes is a practical issue. While there is no theoretical criterion for choosing a proper sample size, in practice we can do correlation/independence simulations on a graph (like in Section 5.2) and choose a large enough sample size so that the recall is above a user defined threshold, e.g. 0.95. Recall is connected to the Type I and Type II errors in statistical tests, for independence and correlation respectively.

Our method can assess correlations in different vicinity levels, i.e. $h$. Another scheme could be that we get rid of $h$ by designing a weighted correlation measure where reference nodes closer to event nodes have higher weights. This is challenging since we cannot directly make use of $\tau$'s nice property of being asymptotically normal in the null case. Another possible extension is to consider event intensity on nodes, e.g. the frequency by which an author used a keyword. We leave these possible extensions for future work.

## 7. RELATED WORK

Our work is related to a branch of graph mining research which involves both graph structures and node attributes [27, 8, 20, 22, 16, 11]. Zhou et al. proposed a graph clustering algorithm based on both structural and attribute similarities [27]. In [8], Ester et al. also investigated using node attribute data to improve graph clustering. Moser et al. introduced the problem of mining cohesive graph patterns which are defined as dense and connected subgraphs that have homogeneous node attribute values [20]. Although the above works considered both graph structures and node attributes, they did not explicitly study the relationships between structures and attributes. Recently, Silva et al. [22] proposed a structural correlation pattern mining problem which aims to find pairs $(S, V)$ where $S$ is a frequent attribute set and $V$ induces a dense subgraph. Each node in $V$ contains all the attributes in $S$. However, this kind of correlation is too restrictive and strong. An attribute which occurs on $|V|-1$ nodes of $V$ will be discarded, though it also has a positive correlation with attributes in $S$. While our approach allows a user to measure the structural correlation between any attributes freely. Kahn et al. studied the problem of mining a set of attributes which frequently co-occurred in local neighborhoods in a graph [16]. They also tried to assess the significance of the discovered patterns. Nevertheless, our problem is significantly different from theirs as shown in Sections 1 and 5.4. In [11], we proposed a measure based on hitting time to assess the structural correlation within an attribute. Significance of the correlation is estimated via a normal approximation of the measure's distribution under the null hypothesis where variances are estimated by simulations. However, this measure is not suitable for **TESC** in that if we adapt the measure to compute the affinity between two events its distribution in the null case is difficult to estimate by simulations. It is hard to preserve each event's internal structure when simulating independence between them.

Our work is also related to assessing and testing the correlation between two spatial point patterns in spatial pattern analysis [7, 18, 23]. However, existing solutions for this similar problem cannot be applied directly to graph spaces due to following reasons: (1) proximity measures for continuous spatial spaces cannot be applied to graph spaces directly; (2) the fixed and discrete graph structure renders infeasible some popular testing methodologies such as randomly shifting one point pattern around the space [18]; (3) focusing on regions where points exist and uniformly sampling reference points are trivial works in continuous spaces [23], while it is not in our problem due to the discrete nature of graphs. (4) in our case, scalability is an important issue, which existing methods for the point pattern problem failed to consider.

## 8. CONCLUSIONS

We studied the problem of measuring Two-Event Structural Correlations (TESC) in graphs and proposed a novel measure and an efficient testing framework to address it. Given the occurrences of two events we choose uniformly a sample of reference nodes from the vicinity of all event nodes and compute for each reference node the densities of the two events in its vicinity respectively. Then we employ the Kendall's $\tau$ rank correlation measure to compute the average concordance of density changes for the two events, over all pairs of reference nodes. Correlation significance can then be assessed by $\tau$'s nice property of being asymptotically normal under the null hypothesis. We also proposed three different algorithms for efficiently sampling reference nodes. Another rank correlation statistic, Spearman's $\rho$ [15] could also be used. We choose Kendall's $\tau$ since it can provide an intuitive interpretation and also facilitate the derivation of the efficient importance sampling method. Finally, experiments on real graph datasets with both synthetic and real events demonstrated that the proposed TESC testing framework was not only efficacious, but also efficient and scalable.

## 9. ACKNOWLEDGMENTS